\documentclass[aps,prb,groupedaddress,superscriptaddress,twocolumn]{revtex4-1}
\usepackage{amsmath,graphicx}

\begin{document}


\title{Thermal dissipation in the quantum Hall regime in graphene}

\author{Jing-Yun Fang}
\affiliation{International Center for Quantum Materials, School of Physics, Peking University, Beijing 100871, China}
\affiliation{CAS Center for Excellence in Topological Quantum Computation, University of Chinese Academy of Sciences, Beijing 100190, China}

\author{Ning-Xuan Yang}
\affiliation{International Center for Quantum Materials, School of Physics, Peking University, Beijing 100871, China}
\affiliation{CAS Center for Excellence in Topological Quantum Computation, University of Chinese Academy of Sciences, Beijing 100190, China}

\author{Qing Yan}
\affiliation{International Center for Quantum Materials, School of Physics, Peking University, Beijing 100871, China}
\affiliation{CAS Center for Excellence in Topological Quantum Computation, University of Chinese Academy of Sciences, Beijing 100190, China}
\author{Ai-Min Guo}
\affiliation{Hunan Key Laboratory for Super-microstructure and Ultrafast Process, School of Physics and Electronics, Central South University, Changsha 410083, China}
\author{Qing-Feng Sun}
\email[]{sunqf@pku.edu.cn}
\affiliation{International Center for Quantum Materials, School of Physics, Peking University, Beijing 100871, China}
\affiliation{CAS Center for Excellence in Topological Quantum Computation, University of Chinese Academy of Sciences, Beijing 100190, China}
\affiliation{Beijing Academy of Quantum Information Sciences, West Bld.\#3, No.10 Xibeiwang East Rd., Haidian District, Beijing 100193, China}

\date{\today}

\begin{abstract}
It is widely accepted that both backscattering and dissipation cannot
occur in topological systems because of the topological protection.
Here we show that the thermal dissipation can occur in the quantum Hall (QH)
regime in graphene in the presence of dissipation sources,
although the Hall plateaus and the zero longitudinal resistance still survive.
Dissipation appears along the downstream chiral
flow direction of the constriction in the Hall plateau regime,
but it occurs mainly in the bulk in the Hall plateau transition regime.
In addition, dissipation processes are accompanied with the evolution of
the energy distribution from non-equilibrium to equilibrium.
This indicates that topology neither prohibits the appearance
of dissipation nor prohibits entropy increasing,
which opens a new topic on the dissipation in topological systems.
\end{abstract}

\maketitle

\section{\label{secS1}Introduction}
Topological systems have been attracting extensive and ongoing interests \cite{Kane1,Kane2,Bernevig,Qi,Hasan,Nayak}, because of potential applications in low-dissipation electronic devices and topologically protected edge or surface states \cite{Qi,Hasan,Nayak}.
The quantum Hall (QH) effect is a prime phenomenon in topological systems \cite{Klitzing1,Buttiker2,Tsui,He1}, which was firstly discovered in a two-dimensional electron gas (2DEG) system \cite{Klitzing1}. With the influence of a strong perpendicular magnetic field, the energy spectrum forms a series of impurity broadened discrete Landau levels, where the extended states exist in center and the localized states exist at the edge of the band \cite{Laughlin1,Laughlin2,Laughlin3}. This special energy band leads to many peculiar properties. For example, it supports topologically protected chiral edge states, which are characterized by the topological properties of the wave functions distribution in two-dimensional lattice momentum space, and the corresponding topological invariant is TKNN number, also known as Chern number \cite{Hasan,NiuQ}.

When the Fermi energy is between the Landau levels, the chiral edge states exhibit quantized Hall resistance and the corresponding longitudinal resistance is zero \cite{Halperin}. The QH effect can occur in macroscopic systems. In experiment, the quantized Hall resistance is very specified and it is insensitive to the details of the sample, which leads to the establishment of a new metrological standard \cite{Delahaye,Jeckelmann}.  Moreover, the charge carriers in the chiral edge states are very resistant to scattering and are expected to used in
dissipationless electronic devices \cite{Buttiker2}.

Graphene, a kind of two-dimensional topological material with an isolated single layer hexagonal lattice of carbon atoms, has been found to be an ideal material for realization of the QH effect recently \cite{Novoselov1,ZhangY,Novoselov2}. Graphene has unique band structure with a linear dispersion relation near the Dirac points, which leads to many fancy properties \cite{Castro,Beenakker,Semenoff}. Its quasiparticles obey the massless Dirac-type equation and its Hall plateaus are assumed at the half-integer values \cite{Gusynin,Novoselov1,ZhangY}.

The dissipation sources (e.g., the electron-phonon interaction, impurities)
inevitably exist in real systems and could lead to
thermal dissipation with the energy transfer from electric energy to heat energy
in charge transport processes~\cite{aref11,aref12}.
As dissipation is usually associated with resistance generating~\cite{WangSW},
one would expect that dissipation cannot occur in the QH plateau regime
albeit the existence of dissipation sources,
and dissipation only appears
in the QH plateau transition regime.
However, it was shown that dissipation can take place
at the QH edges through resonant impurities and electron-phonon interactions \cite{ZhangG,Slizovskiy}.
In particular, using high-sensitive non-contact
nano-thermometer \cite{Eriksson1,Halbertal1,Halbertal2},
Marguerite \textit{et al.} \cite{Marguerite} reported an amazing experiment
that dissipative transport happens along the chiral QH edge states in graphene
but in the bulk no dissipation occurs in either QH plateau
or plateau transition regimes.
In view of this, a thorough and reliable analysis of thermal dissipation
accompanied with energy relaxation and electron redistribution
is urgent in the QH effect.

In this paper, we theoretically study the thermal dissipation
of a six-terminal graphene device under a perpendicular magnetic field $B$.
The thermal dissipation processes require energy transfer from
electronic system to environment.
Here we simulate the dissipation sources by introducing
the B\"{u}ttiker's virtual leads \cite{Buttiker1},
where the energy of moving electrons leaks into
these virtual leads and induces thermal dissipation.
By using the tight-binding model and the Landauer-B\"{u}ttiker formalism
together with the non-equilibrium Green's function,
the local heat generation and equivalent temperature are calculated.
Our results indicate that the thermal dissipation
can occur in the QH plateau regime,
deviating from the principle that backscattering and dissipation cannot
happen in the QH effect due to the topological protection.
The thermal dissipation mainly appears along the downstream chiral flow
direction of the constriction, with the relaxation length
determined by the dissipation strength.
These features are in excellent agreement with the recent experiment \cite{Marguerite}.
On the other hand, in the plateau transition regime, the
thermal dissipation primarily appears in the bulk of graphene,
which is the same as previous physical intuition \cite{Buttiker2,Marguerite}.

The rest of the paper is organized as follows. In Sec.~\uppercase\expandafter{\romannumeral2}, we describe the model of the system and give the details of our calculations. In Sec.~\uppercase\expandafter{\romannumeral3}, we show the numerical results and some discussions. Finally, a brief conclusion is presented in Sec.~\uppercase\expandafter{\romannumeral4}.

\begin{figure}
\includegraphics[scale=0.33]{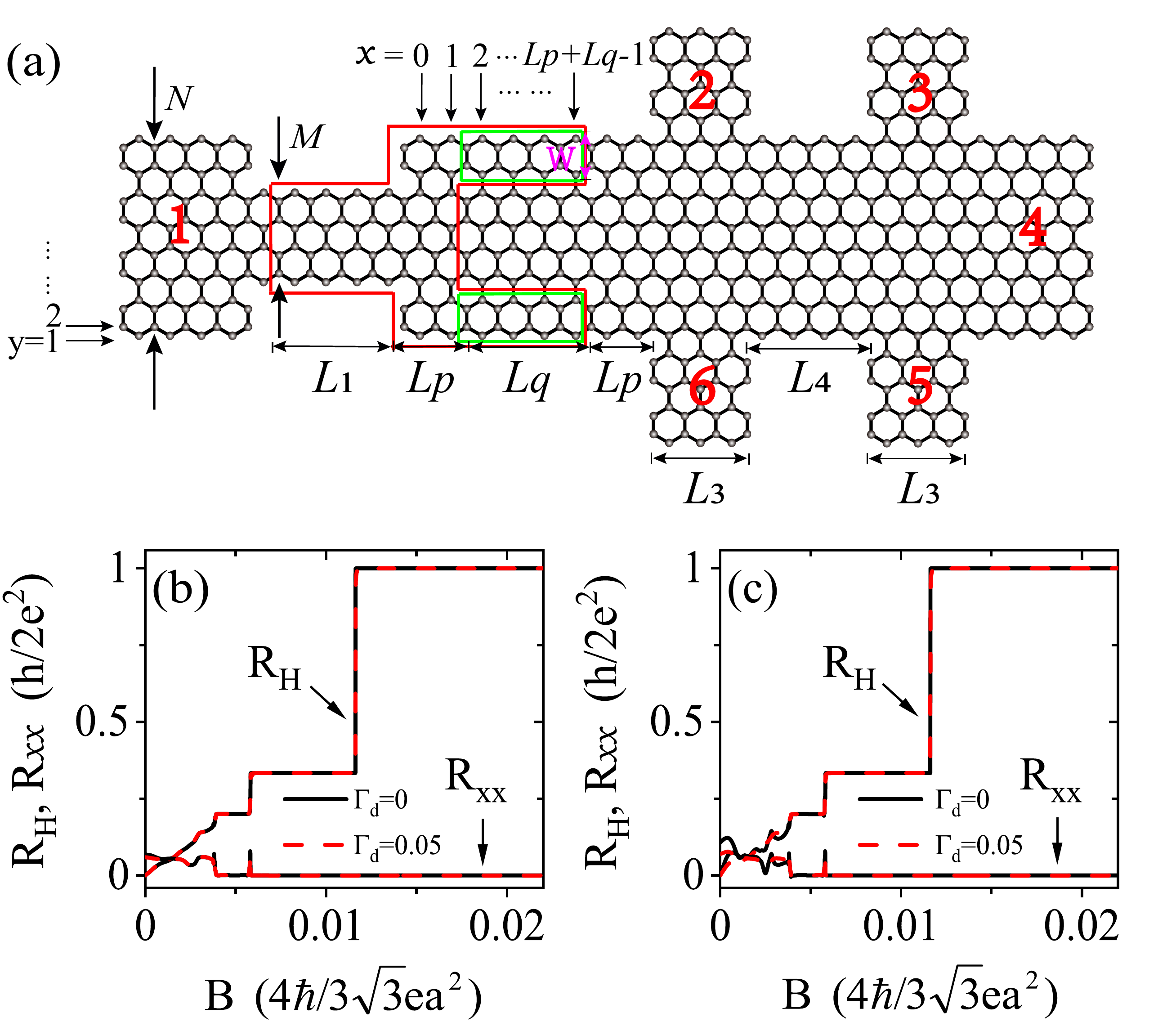}
\centering
\caption{ (a)
Schematic diagram for a six-terminal graphene device with a narrow constriction on the left side (i.e., the narrow constriction is between lead 1 and leads 2, 6).
The region surrounded by the red (green) solid lines contains dissipation sources.
In this diagram, the sizes of device are $N=16$, $M=8$, $W=4$,
$L_{1}=4$, $L_{p}=2$, $L_{q}=4$, $L_{3}=3$ and $L_{4}=4$.
(b) and (c) show the Hall resistance $R_{H}$ and longitudinal resistance $R_{xx}$
vs the magnetic field $B$ with different
dissipation strengths $\Gamma_{d}$ for $M=N$ and $M=8$, respectively.
$L_{q}=80$ and the dissipation sources are in the red solid lines surrounding region.}
\label {fig:1}
\end{figure}

\section{\label{secS2}MODEL AND METHOD}
In the tight-binding representation, the Hamiltonian of a six-terminal graphene device
[see Fig.~\ref{fig:1}(a)] can be written as \cite{Sheng,Suna,Long,Li}
\begin{eqnarray}\label{eq:1}
H&=&\sum_{\bf i}\epsilon_{\bf i}c_{\bf i}^{\dag}c_{\bf i}-\sum_{\langle{\bf ij}\rangle}
 te^{i\phi_{\bf ij}}c_{\bf i}^{\dag}c_{\bf j}
+H_d.
\end{eqnarray}
The first and second terms describe the graphene system
including central region and six terminals.
$c_{\bf i}^\dag$ ($c_{\bf i}$) is the creation (annihilation) operator
at site ${\bf i}$, $\epsilon_{\bf i}$ is the on-site energy,
and $t$ is the nearest-neighbor coupling.
Here ${\bf i}=(x,y)$ is also the spatial position coordinate, as shown in Fig.~\ref{fig:1}(a).
The magnetic field $B$ is expressed as
$\phi_{\bf ij}=\int_{\bf i}^{\bf j}{\vec{A}\cdot{d\vec{l}}/\phi_{0}}$,
with the vector potential $\vec{A}=(-By,0,0)$ in leads 1, 4 and the central region, $\vec{A}=(0,Bx,0)$ in leads 2, 3, 5, 6, and $\phi_{0}=\hbar/e$ the flux quantum, see the Appendix A.
We stress that our results still hold in square-lattice systems (or 2DEG systems).
The third term
$H_d = \sum_{{\bf i},k}\epsilon_{k}a_{{\bf i}k}^{\dag}a_{{\bf i}k}
 +(t_{k}a_{{\bf i}k}^{\dag}c_{\bf i}+\textrm{H.c.})$ represents
 the Hamiltonian of B\"{u}ttiker's virtual leads and their couplings to central sites,
 which is used to simulate the dissipation sources.
$a_{{\bf i}k}^\dag$ ($a_{{\bf i}k}$) is the creation (annihilation) operator of the electrons in the virtual lead ${\bf i}$,
$t_{k}$ is the coupling strength between the virtual leads and the graphene.
In fact, the virtual leads can be used as dephasing probes, voltage probes or temperature probes,
which depend on the boundary conditions of the virtual leads.
When the virtual leads are set as dephasing probes,
electrons can lose phase memories by going into and coming back from the virtual leads,
and the electric currents in these virtual leads are zero~\cite{XingY}.
When they are set as voltage probes, the electric currents in the virtual leads are also zero
and the voltage in the virtual leads are studied.
While when they are set as temperature probes, both the heat currents and the electric currents are zero.
Here we use the virtual leads to simulate the dissipation sources,
in this case electrons can lose energy by going into and coming back from the virtual leads,
i.e., the electric currents are zero and we focus on the heat currents in the virtual leads.
In addition,  we consider that the dissipation sources only exist
in the region enclosed by the red (green) solid lines in Fig.~\ref{fig:1}(a),
where each site is attached by a virtual lead.

By using the multiprobe Landauer-B\"{u}ttiker formula,
the electric current and the heat current in real leads $r$ ($r=1$, $2$,..., $6$) and virtual ones ${\bf i}$
can be expressed as \cite{YangNX,Datta,Sancho}
\begin{eqnarray}\label{eq:2}	
J_{p}&=& \frac{2e}{h}\sum_{q}\int{T_{pq}(E)[f_{p}(E)-f_{q}(E)]dE}, \nonumber  \\
Q_{p}&=&-\frac{2}{h}\sum_{q}\int{(E-\mu_{p})T_{pq}(E)[f_{p}(E)-f_{q}(E)]dE}, \notag \\
\end{eqnarray}
where $p$, $q\in{r}$ or ${\bf i}$.
For convenience, we define the electric current $J_{p}$ flows from lead $p$
into central region as the positive direction,
while the heat current $Q_{p}$ flows from central region to lead $p$ as the positive direction.
Here, the heat current $Q_{p}$ is induced by the flow of the electric current,
so we will call it the current-induced local heat generation in the following.
In Eq.~(\ref{eq:2}), $T_{pq}(E)=\textmd{Tr}[{\bf \Gamma}_{p}{\bf G}^r{\bf \Gamma}_{q}{\bf G}^a]$ is the transmission coefficient
from lead $q$ to lead $p$, where the Green's function ${\bf G}^r(E)=[{\bf G}^a(E)]^\dag=[E {\bf I}-{\bf  H}_{cen}-\sum_{p}{\bf \Sigma}_{p}^r]^{-1}$
and the linewidth function
${\bf \Gamma}_{p}(E)=i({\bf \Sigma}_{p}^r(E)-{\bf \Sigma}_{p}^{a}(E))$.
${\bf H}_{cen}$ is the Hamiltonian of the central region.
${\bf \Sigma}_{p}^r(E) =[{\bf \Sigma}_{p}^{a}(E)]^\dagger$ is the retarded self-energy
due to the coupling to lead $p$.
For real leads $p$ ($p\in{r}$), the self-energy ${\bf \Sigma}_{p}^r$ can be calculated
numerically \cite{Sancho}.
For virtual leads $p$ ($p\in{\bf i}$), ${\bf \Sigma}_{p}^r=-\frac{i}{2} \Gamma_{d}$,
with the dissipation strength $\Gamma_{d} =2\pi\rho t_k^2$ and
$\rho$ the density of states in the virtual leads~\cite{XingY}.
We assume that $\Gamma_{d}$ is independent of the energy, i.e., in the wide-band approximation.
$f_{p}(E)=[e^{(E-\mu_{p})/k_B\mathcal{T}_{p}}+1]^{-1} $
is the Fermi distribution function in lead $p$,
with the temperature $\mathcal{T}_{p}$,
the chemical potential $\mu_{p}=E_{F}+eV_{p}$, and the voltage $V_{p}$ in lead $p$.

\begin{figure}[htbp]
	\includegraphics[scale=0.37]{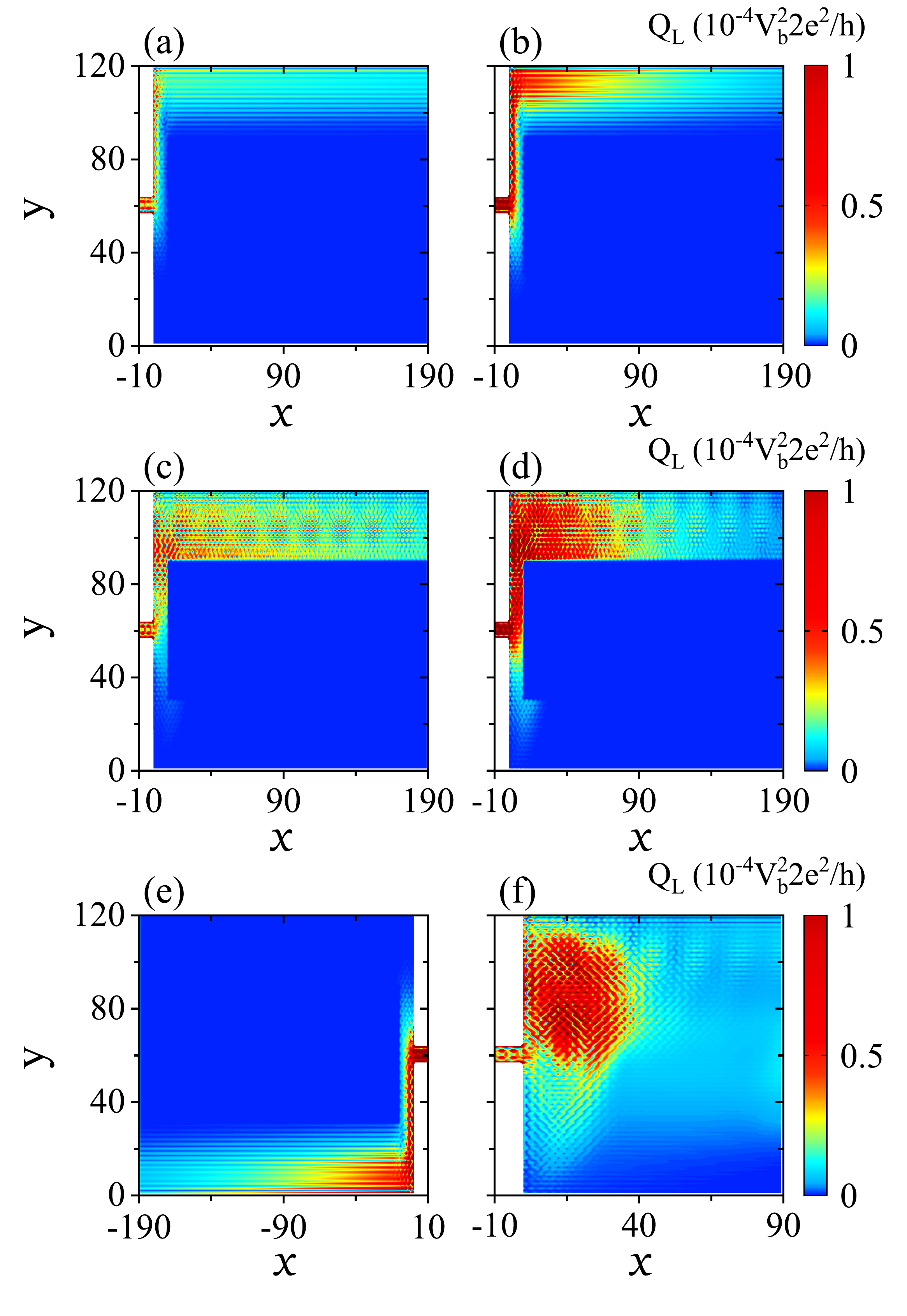}
	\centering
	\caption{
		(a-d) show the local heat generation $Q_L$ vs lattice position $(x,y)$ in the QH plateau regime with a narrow constriction on the left side (i.e., the narrow constriction is between lead 1 and leads 2, 6 in Fig.~\ref{fig:1}(a)). $B=0.02$ ($\nu=1$) in (a), (b), $0.01$ ($\nu=3$) in (c), (d), and dissipation strengths $\Gamma_{d}=0.005$ in (a), (c), $0.02$ in (b), (d). (e) shows $Q_L$ at $B=0.02$ for the device with
		a narrow constriction on the right side (i.e., the narrow constriction is between lead 4 and leads 3, 5 in Fig.~\ref{fig:1}(a)) and $\Gamma_{d}=0.02$.
		(f) shows $Q_L$ in the Hall plateau transition regime ($B=0.0116$) with
		$L_{q}=80$, $W=60$ and $\Gamma_{d}=0.005$.
		The dissipation sources are in the region surrounded by the red solid lines
		in Fig.~\ref{fig:1}(a).
	}
	\label {fig:2}
\end{figure}

By using the Green's function, the energy distribution function at the site ${\bf i}$ can be expressed as
\begin{eqnarray}\label{eq:3}
F_{\bf i}(E)=\frac{n({\bf i},E)}{LDOS_{\bf i}(E)},
\end{eqnarray}
where $n({\bf i},E)=-\frac{i}{2\pi}{\bf G}_{\bf ii}^<(E)$
is the electron density per unit energy at the lattice site ${\bf i}$
and $LDOS_{\bf i}(E)=-\frac{1}{\pi}\textmd{Im}{\bf G}_{\bf ii}^{r}(E)$
is the local density of states at the lattice site ${\bf i}$~\cite{Datta}.
According to the Keldysh equation, the lesser Green's function
${\bf G}_{\bf ii}^{<}(E)=\sum_{p}{\bf G}_{{\bf i}p}^{r}{\bf \Sigma}_{p}^{<}{\bf G}_{p{\bf i}}^{a}$ and the lesser self-energy ${\bf \Sigma}_{p}^{<}(E)=-f_{p}(E)({\bf \Sigma}_{p}^{r}(E)-{\bf \Sigma}_{p}^{a}(E))$.

When a small bias $V_{b}$ is applied between leads 1 and 4 with $V_1=V_b$ and $V_4=0$,
the current flows along the longitudinal direction.
It is reasonable to assume that the transmission coefficient $T_{pq}(E)$
is approximately independent of energy $E$ since the bias is small.
The leads 2, 3, 5, 6 are set as voltage probes and their electric currents are zero.
Besides, the electric currents in the virtual leads are also zero,
because electrons go into and come back from the virtual leads, only losing energy.
The Fermi distribution function is $f_{p}(E)=\theta(\mu_{p}-E)$ at zero temperature.
Thus, Eq.~(\ref{eq:2}) can be expressed as
\begin{eqnarray}\label{eq:4}
J_{p}&=& \frac{2e^2}{h}\sum_{q\ne p}T_{pq}(E_{F})(V_{p}-V_{q}), \nonumber  \\
Q_{p}&=&-\frac{2e^2}{h}\sum_{q\ne p}T_{pq}(E_{F})(V_{p}V_{q}-\frac{1}{2}V_{p}^2-\frac{1}{2}V_{q}^2).
\end{eqnarray}
Combining $J_p=0$ ($p\in\{{\bf i}$, 2, 3, 5, 6\}) with Eq.~(\ref{eq:4}),
the voltages $V_{p}$, the current-induced local heat generation $Q_{p}$ in these leads
and the longitudinal current ($J=J_{1}=-J_{4}$) can be obtained.
The current $J$ is proportional to the bias $V_b$ and
the heat generation $Q_{p}$ to $V_b^2$.
The local heat generation $Q_L(x,y)$ at position ${\bf i}=(x,y)$
is $Q_L(x,y) = Q_{\bf i}$.
Finally, the longitudinal resistance $R_{xx}=R_{14,23}=(V_{2}-V_{3})/J$
and the Hall resistance $R_{H}=R_{14,26}=(V_{2}-V_{6})/J$ can be calculated straightforwardly.
And the local heat generation $Q_L(x,y)$ will be studied in Sec.~\uppercase\expandafter{\romannumeral3}.A.

On the other hand, if the thermal conductivity between the sample and environment is poor,
the local heat generation $Q_{\bf i}$ disappears and the local electron temperature $\mathcal{T}_{\bf i}$ 
(i.e., the temperature in the virtual leads) rises.
In this case, the virtual leads act as both the dissipation sources
and the temperature detection terminals (temperature probes),
and both $J_{\bf i}$ and $Q_{\bf i}$ are zero. 
Notice that although both $J_{\bf i}$ and $Q_{\bf i}$ are zero in the virtual leads, the heat current density $\sum_{q} (E-\mu_{p})T_{pq}(E)[f_{p}(E)-f_{q}(E)]$ and
the electric current density $\sum_{q} T_{pq}(E)[f_{p}(E)-f_{q}(E)]$ are usually non-zero.
So the dissipation can occur and the virtual leads still play the role of the dissipation sources.
Provided that the six real leads have the same temperature ($\mathcal{T}_1=\mathcal{T}_2=\mathcal{T}_3
=\mathcal{T}_4=\mathcal{T}_5=\mathcal{T}_6=\mathcal{T}$).
At low temperature and small voltage,
the current $J_{p}$ and the current-induced local heat generation $Q_{p}$
in Eq.~(\ref{eq:2}) can be reduced as
\begin{eqnarray}\label{eq:5}
J_{p}&=&\frac{2e^2}{h}\sum_{q\ne p}T_{pq}(E_{F})(V_{p}-V_{q}), \nonumber   \\
Q_{p}&=&-\frac{2}{h}\sum_{q\ne p}
T_{pq}(E_{F})\left[\frac{\pi^{2}}{6}k_{B}^2(\mathcal{T}_{p}^2-\mathcal{T}_{q}^2)
 -\frac{1}{2}e^2(V_{p}-V_{q})^2\right]. \notag \\
\end{eqnarray}
Here $k_{B}$ is the Boltzmann constant and
$\mathcal{T}_{p}$ is the temperature in lead $p$.
In the expression of the current-induced local heat generation $Q_p$ in Eq.~(\ref{eq:5}),
the linear term $\Delta \mathcal{T}_{p} \equiv \mathcal{T}_{p} -\mathcal{T}$ exists,
but the linear term $V_{p}-V_{q}$ disappears
because we now focus on the heat generation
at the low temperature and small voltage, see the Appendix B for the detail.
This is different from some previous literatures on
thermoelectric effects and Peltier effect in linear response,
which includes the linear term $V_{p}-V_{q}$~\cite{addTE1,addTE2}.
Notice that the temperature $\mathcal{T}_{p}$ ($p\in {\bf i}$) in the virtual leads
may not be equal to the background temperature $\mathcal{T}$.
Then by using the boundary conditions that the net currents $J_{p}$
flowing through real leads 2, 3, 5, 6
and all the virtual leads are zero
and the heat generation $Q_{p}$ in all the virtual leads are also zero,
the local equivalent electron temperature $\mathcal{T}_{\bf i}$
and voltage $V_{p}$ ($p\in r$ or ${\bf i}$) can be obtained.
Also note that the electron at the site ${\bf i}$ of the graphene is in non-equilibrium.
Here $\mathcal{T}_{\bf i}$ is the temperature of the virtual lead ${\bf i}$, which is in equilibrium.
Because that the virtual leads are in thermal contact with graphene and
the heat current $Q_{\bf i}$ from the graphene to the virtual leads is zero,
we can use the temperature $\mathcal{T}_{\bf i}$ as an equivalent temperature of the local
non-equilibrium electron at the site ${\bf i}$.
In Sec.~\uppercase\expandafter{\romannumeral3}.B, we will numerically study
the local equivalent temperature $\mathcal{T}_{\bf i}$.

\section{\label{secS3}NUMERICAL RESULTS AND DISCUSSIONS}

In the numerical calculations, we set the hopping energy $t=2.75$ eV,
the on-site energy $\epsilon_{\bf i}=0.2t$ (i.e., the energy of Dirac point), and the Fermi energy $E_{F}=0$. Taking into account the spin degeneracy, we will use $h/2e^{2}$ as the resistance unit. The corresponding filling factors are taken as odd integers ($\nu=1$, 3, 5...) instead of even integers ($\nu=2$, 6, 10...), which are actually in accordance with the experiment result \cite{Marguerite}.
The zigzag edge graphene ribbon is considered [Fig.~\ref{fig:1}(a)] and the results still hold for the armchair one.
The magnetic field is expressed
as the Peierls phase \cite{Long,Cresti}: $2\phi=(3\sqrt{3}/2)a^2B/\phi_{0}$,
with $(3\sqrt{3}/2)a^2B$ the magnetic flux threading a single hexagon and the unit of $B$ being $4\hbar/(3\sqrt{3}ea^2)$, where $a=0.142$ nm is the lattice constant of graphene.
The device sizes are $N=120$, $W=30$, $L_{1}=10$, $L_{p}=10$, $L_{q}=180$,
$L_{3}=50$, $L_{4}=50$, $B=0.02$, and the temperature $\mathcal{T}=0$ for all leads, which will be used in the calculations unless stated otherwise.

\begin{figure}
	\includegraphics[scale=0.37]{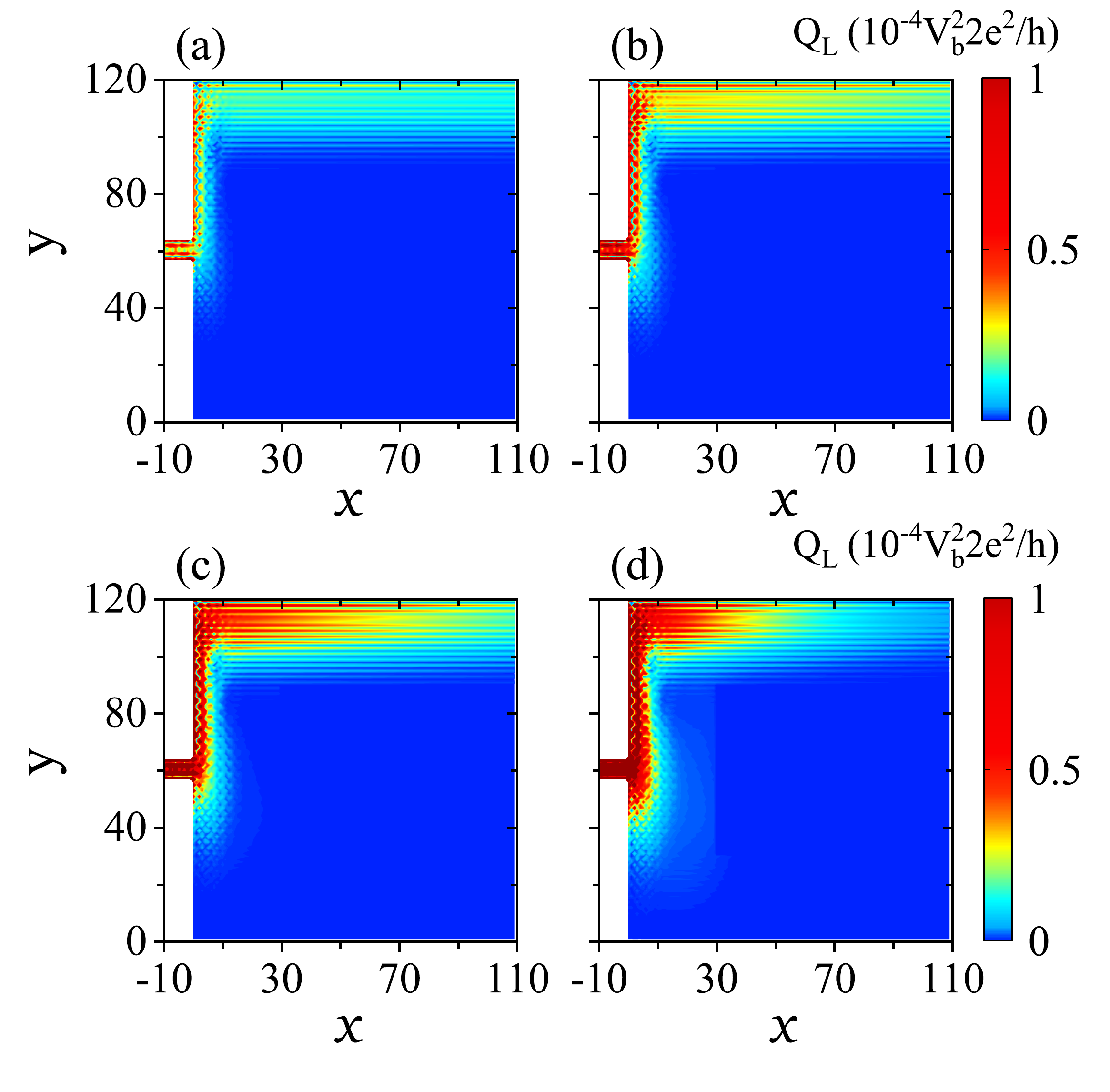}
	\centering
	\caption{Local heat generation $Q_{L}$
		vs lattice position $(x,y)$ with $L_{p}=30$, $L_{q}=80$. The dissipation sources are
		in the region surrounded by the red solid lines [see Fig.~\ref{fig:1}(a)]
		and dissipation strength $\Gamma_d=0.005$ (a), $0.01$ (b), $0.02$ (c) and $0.05$ (d).}
	\label{fig:3}
\end{figure}

\subsection{\label{A}Hall resistance and thermal dissipation}

Figures~\ref{fig:1}(b) and \ref{fig:1}(c) show the Hall resistance $R_{H}$
and the longitudinal one $R_{xx}$
as functions of magnetic field $B$ with different dissipation strengths $\Gamma_{d}$.
In the absence of dissipation sources ($\Gamma_d=0$) and narrow constriction,
$R_H$ increases with increasing $B$ and $R_{xx}$ oscillates.
At large $B$, the expected plateaus at $R_H=(1/\nu)h/2e^{2}$ with filling factors $\nu=1$, 3, 5... are found, and
$R_{xx}$ is zero except in the plateau transition regime,
owing to the formation of Landau levels and chiral edge states \cite{Datta,Giamarchi}.
In particular, the QH plateaus and zero $R_{xx}$ remain well
in the case of either dissipation sources or narrow constriction exist.
These phenomena are consistent with previous experimental \cite{Klitzing1,Novoselov1,ZhangY,Novoselov2,Delahaye}
and theoretical works \cite{Datta,Buttiker2,XingY,Jiang}.
That is, the QH effect is topologically protected,
so no dissipation and no backscattering are naively expected.

Next, we focus on the heat generation in the presence of the dissipation sources.
Without the narrow constriction, i.e., $M=N=120$,
the current-induced local heat generation $Q_{L}$
is almost zero in the QH plateau regime,
because the charge carriers flow along the topologically protected chiral edge states.
This means that thermal dissipation and backscattering cannot occur as expected.
However, the results are different in the presence of narrow constriction ($M=8$).

Figures~\ref{fig:2}(a) and \ref{fig:2}(b) plot
the local heat generation $Q_{L}$
at $M=8$ in the QH plateau regime with $B=0.02$ ($\nu=1$). When $B=0.02$, the magnetic length $l_{B}=\sqrt{\frac{\hbar}{eB}}\approx8a \sim M$,
thus backscattering will happen at the constriction
and the electrons will be in non-equilibrium states after passing through the constriction,
leading to voltage drop and work generation at the constriction.
From Figs.~\ref{fig:2}(a) and \ref{fig:2}(b),
one can see that $Q_{L}$ is quite large at the constriction region and the device edges,
indicating the emergence of thermal dissipation, contrary to the intuition that the QH effect is dissipationless.
Furthermore, thermal dissipation mainly appears along
the downstream chiral flow direction of the constriction and is very weak in the bulk.
These results are in good agreement with the recent experiment \cite{Marguerite}.
When away from the constriction, the heat generation $Q_{L}$ is gradually declined
with a relaxation length $\lambda$.
For small dissipation strength $\Gamma_d$, $\lambda$ is very long and
$Q_{L}$ is almost the same along the downstream channel [Fig.~\ref{fig:2}(a)],
similar to the experimental results \cite{Marguerite}.
While for large $\Gamma_d$, $\lambda$ is short and $Q_L$ decays significantly [Fig.~\ref{fig:2}(b)].
With the increase of dissipation strength $\Gamma_{d}$,
the total local heat generation $Q_{LT}=\sum_{x,y}Q_{L}(x,y)$ in the virtual leads increases.
Notice that although the thermal dissipation appears,
the QH plateaus and zero longitudinal resistance $R_{xx}$ remain well [Figs.~\ref{fig:1}(b) and \ref{fig:1}(c)].

Figures~\ref{fig:2}(c) and \ref{fig:2}(d) plot
the local heat generation $Q_{L}$ for the higher filling factor $\nu=3$ ($B=0.01$).
For the higher $\nu$, the thermal dissipation increases significantly, since there are more edge states.
From Figs.~\ref{fig:2}(c) and \ref{fig:2}(d),
we can see that $Q_{L}$ mainly appears along the downstream chiral flow direction of the constriction,
and is still very weak in the bulk.
When away from the constriction, the heat generation $Q_{L}$ is gradually declined with a relaxation length $\lambda$.
With the increase of dissipation strength $\Gamma_{d}$,
the total local heat generation $Q_{LT}$ increases
while the relaxation length $\lambda$ decreases.
These results are similar to the case of filling factor $\nu=1$.
In addition, for the higher filling factor,
the thermal dissipation is slightly delocalized to the edge of the system,
because the higher edge states are more extended.
Figure~\ref{fig:2}(e) shows the local heat generation $Q_{L}$ for the situation that
the constriction locates at the right side.
Now the thermal dissipation mainly occurs at the lower edge,
which is still aligned with the downstream chiral flow direction of the constriction.

\begin{figure}
	\includegraphics[scale=0.37]{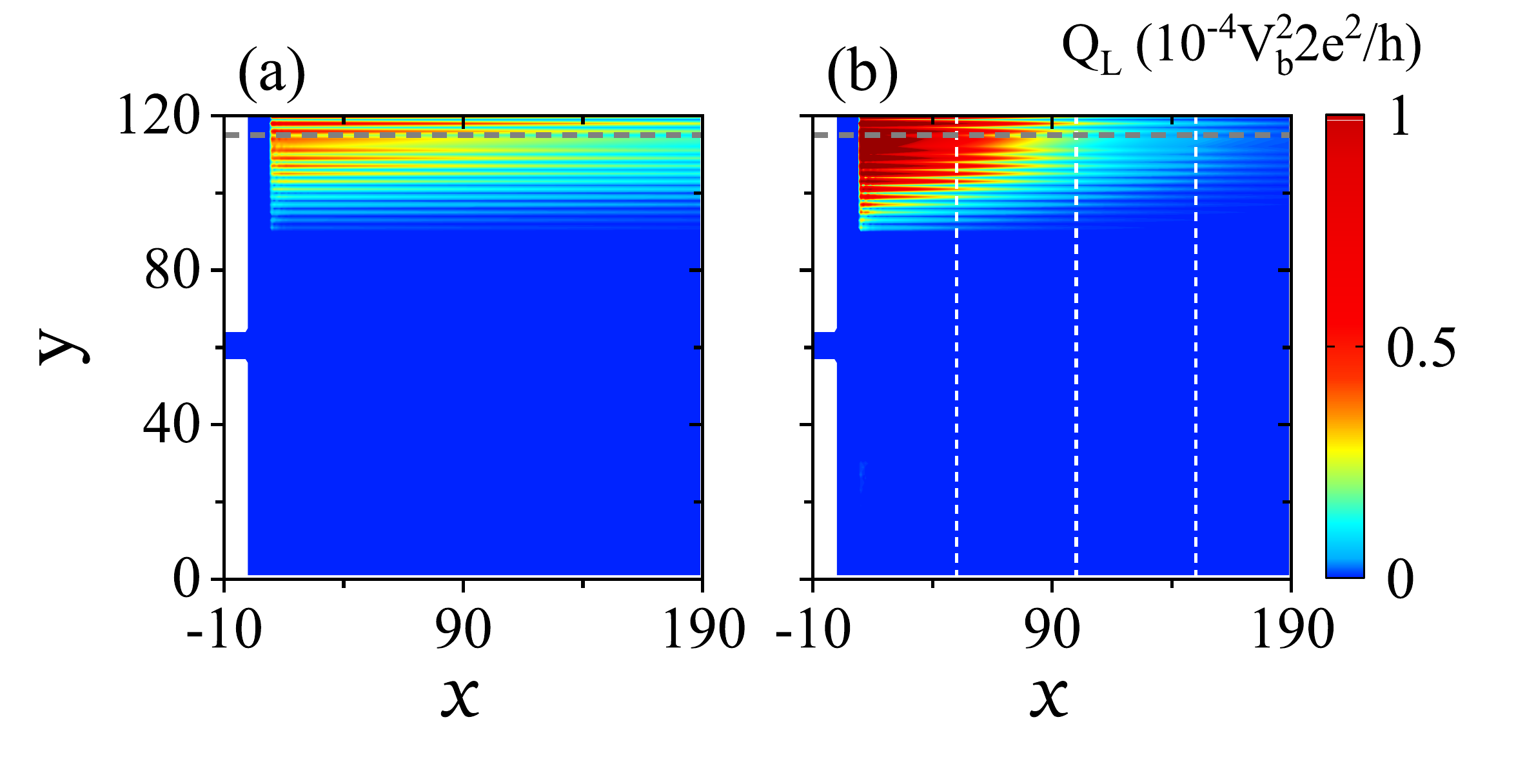}
	\centering
	\caption{The local heat generation $Q_{L}$ vs lattice position $(x,y)$
		for dissipation strengths $\Gamma_{d}=0.01$ in (a) and $\Gamma_{d}=0.05$ in (b).
		The dissipation sources are
		in the region surrounded by the green solid lines [see Fig.~\ref{fig:1}(a)].
		$Q_{L}$ along the white or gray dotted lines in (a) and (b)
		are shown in Figs.~\ref{fig:5}(a) and \ref{fig:5}(b). }
	\label {fig:4}
\end{figure}

While in the Hall plateau transition regime,
one can see from Fig.~\ref{fig:2}(f) that the thermal dissipation mainly occurs
in the bulk as expected, because the Fermi energy locates at the spatially extended Landau level.
The thermal dissipation is slightly larger in the upper part than
the lower one, as the electrons move toward the upper part under
the magnetic field, which is different from the recent experiment
where the thermal dissipation occurs along both the downstream and upstream directions with no visible chirality \cite{Marguerite}.
This experimental phenomenon may originate from the edge reconstruction
in graphene \cite{Silvestrov,Cui,reconstr1},
which induces additional non-topological
counterpropagating channels \cite{NatP.Uri}.
Our results predict that the thermal dissipation mainly
occurs in the bulk in the plateau transition regime
if there is no edge reconstruction in graphene
or for 2DEG systems (e.g., GaAs/AlGaAs heterostructures).
It is worth mentioning that the edge reconstruction
has little effect on the thermal dissipation in the QH plateau regime,
because the Fermi energy $E_F$ locates far from the Landau levels in this case.

\begin{figure}
	\includegraphics[scale=0.3]{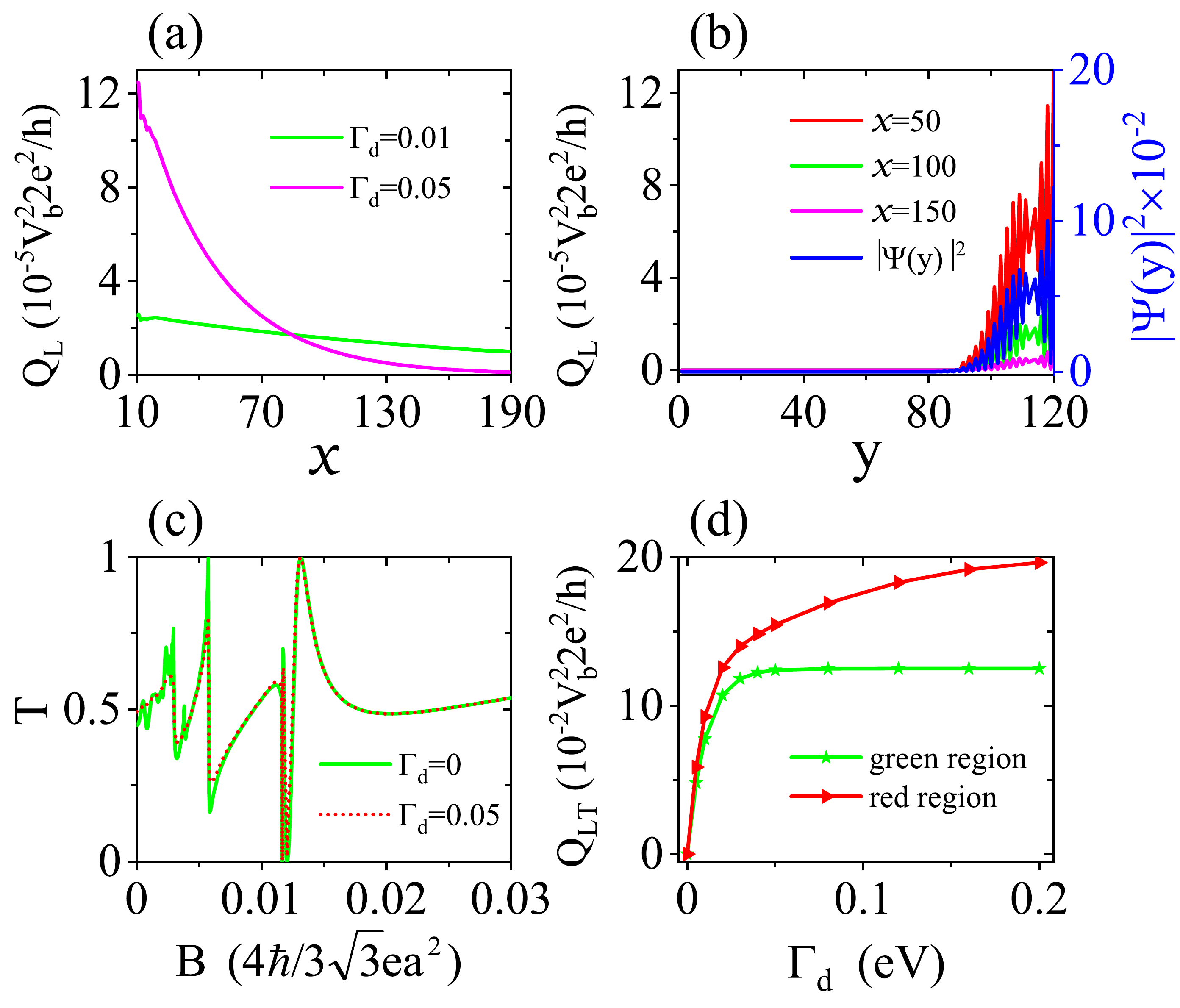}
	\centering
	\caption{(a) The local heat generation $Q_L$ vs longitudinal location $x$ under fixed $y=115$ for different $\Gamma_d$. (b) shows $Q_L$ vs transverse location $y$ with $\Gamma_d=0.05$  for different $x$,
		and the wave function $|\Psi(y)|^2$ vs location $y$.
		(c) The transmission coefficient $T$ of the narrow constriction
		vs the magnetic field $B$ for different $\Gamma_d$.
		(d) shows the total local heat generation $Q_{LT}$
		versus dissipation strength $\Gamma_{d}$. In (a-c), the dissipation sources exist
		in the region surrounded by the green solid lines in Fig.~\ref{fig:1}(a).
		In (d), the red and green curves correspond to
		the dissipation sources in the regions enclosed by the red and green
		solid lines, respectively.}
	\label {fig:5}
\end{figure}

Now, we offer a detailed discussion that $L_{p}=10$ in Fig.~\ref{fig:2} is reasonable
and the local heat generation is almost zero in the bulk in the Hall plateau regime.
Figure~\ref{fig:3} shows the local heat generation $Q_L$ for $L_p=30$
and the magnetic field $B=0.02$. When $B=0.02$,
the system is in the first QH plateau regime.
From Fig.~\ref{fig:3}, we can see that for the case of $L_{p}=30$, which is much larger than $l_{B}$, the
thermal dissipation along the left edge still occurs, and mainly appears
in the region with $x<10$, regardless of the dissipation strength $\Gamma_{d}$.
Therefore, it is reasonable to assume $L_{p}=10$.
What's more, there is no thermal
dissipation in the bulk in the Hall plateau regime.
The thermal dissipation mainly occurs along the downstream chiral flow direction
of the narrow constriction, with a relaxation length $\lambda$
related to the dissipation strength.
With the increase of dissipation strength $\Gamma_{d}$,
the total local heat generation $Q_{LT}$ in the virtual leads increases,
while the relaxation length reduces.
These results are completely the same as those in Figs.~\ref{fig:2}(a) and \ref{fig:2}(b).

In the discussion above, the dissipation sources exist at the constriction,
so that the transmission coefficient of the constriction will be affected by $\Gamma_d$.
In order to eliminate the interaction between the constriction and the dissipation sources, here we consider the situation that the dissipation
sources only appear in the region enclosed by the green solid lines in Fig.~\ref{fig:1}(a). From Figs.~\ref{fig:4}(a) and \ref{fig:4}(b), we can see a large heat generation $Q_{L}$ still appears
along the downstream chiral flow direction (the upper edge of the system) when $\Gamma_d\not=0$, and the heat generation is almost zero along the upstream chiral flow direction
(the lower edge of the system), just like Figs.~\ref{fig:2}(a) and \ref{fig:2}(b).

\begin{figure}
	\includegraphics[scale=0.37]{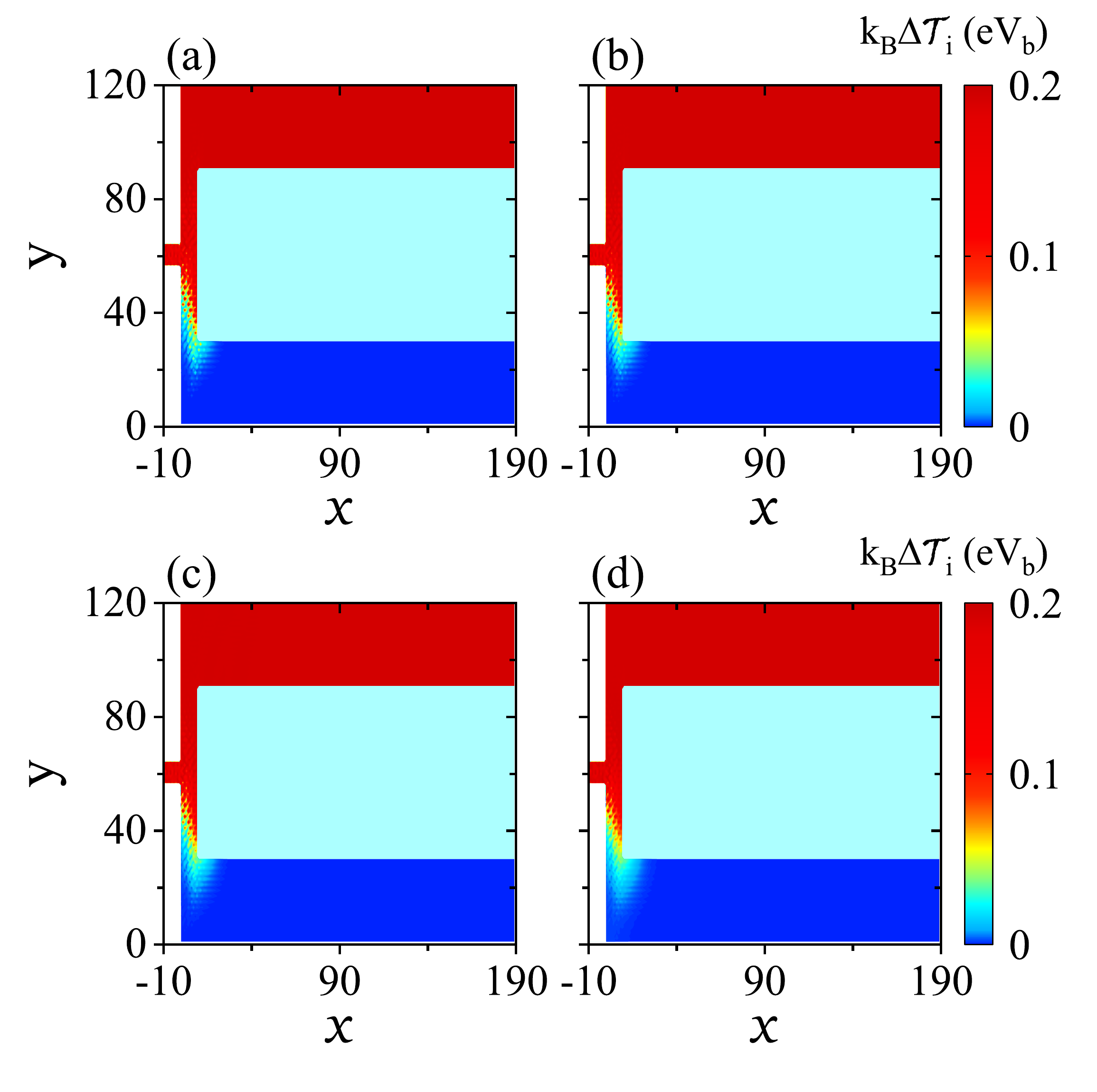}
	\centering
	\caption{Equivalent temperature $k_{B}\Delta\mathcal{T}_{\bf i}$
		($\Delta\mathcal{T}_{\bf i} \equiv \mathcal{T}_{\bf i}-\mathcal{T}$)
		vs lattice position ${\bf i}=(x,y)$ with
		$\Gamma_{d}=0.005$ (a), $0.01$ (b), $0.02$ (c) and $0.05$ (d).
		The background temperature $k_{B}\mathcal{T}/(eV_{b})=0.1$.
		The dissipation sources exist in the region surrounded
		by the red solid lines in Fig.~\ref{fig:1}(a).
		The light cyan area is added artificially to show the part without
		coupling of the virtual leads.}
	\label{fig:6}
\end{figure}

Then, we study the thermal dissipation in more details in the QH plateau regime.
The curves in Figs.~\ref{fig:5}(a) and \ref{fig:5}(b) are extracted from Fig.~\ref{fig:4} by fixing the transverse location $y$ or the longitudinal location $x$.
Figure~\ref{fig:5}(a) shows the local heat generation $Q_{L}$ versus longitudinal location $x$ with a fixed transverse location $y$ for different $\Gamma_{d}$.
For small dissipation strength $\Gamma_{d}$,
the local heat generation $Q_{L}$ is almost the same regardless of $x$.
While for large $\Gamma_{d}$, $Q_{L}$ is dramatically declined with increasing $x$.
Figure~\ref{fig:5}(b) plots the local heat generation $Q_{L}$ versus transverse location $y$.
It clearly appears that the thermal dissipation
is almost zero at the lower boundary (the upstream direction)
and mainly occurs at the upper boundary (the downstream direction). Furthermore, the thermal dissipation obviously reduces with the increase of longitudinal location $x$.
For example, $Q_L$ at $x=50$ is much larger than that at $x=100$,
and $Q_{L}$ at $x=150$ is nearly zero.
Besides, at a fixed $x$, the thermal dissipation oscillates with $y$.
The wave function $|\Psi(y)|^2$ of the chiral edge state is also shown in Fig.~\ref{fig:5}(b).
$|\Psi(y)|^2$ oscillates with $y$, which is very similar to the curve $Q_{L}$-$y$.
This indicates that the thermal dissipation originates from
the topologically protected chiral edge states.

When the dissipation sources only appear in the region enclosed by the green solid lines in Fig.~\ref{fig:1}(a), the transmission coefficient through the narrow constriction is $T=\sum_{q=2}^6 T_{q1} +\sum_{\bf i} T_{{\bf i}1}$.
Figure~\ref{fig:5}(c) shows the transmission coefficient $T$ versus the magnetic field $B$.
$T$ is non-integer and depends on the magnetic field $B$.
For $B=0.02$ (in the first Hall plateau regime), $T=0.4858$,
and for $B=0.0116$ (in the Hall plateau transition regime), $T=0.5466$.
In addition, the transmission coefficient $T$ will be hardly affected by the dissipation strength $\Gamma_d$.  Figure~\ref{fig:5}(d) plots the total local heat generation $Q_{LT}$
versus the dissipation strength $\Gamma_{d}$.
Without the dissipation sources ($\Gamma_d =0$),
the total local heat generation $Q_{LT}$ is zero.
With the increase of $\Gamma_d$, $Q_{LT}$ increases monotonically.
For the dissipation sources only exist at the sample
(in the green solid lines surrounding region in Fig.~\ref{fig:1}(a)),
$Q_{LT}$ at large $\Gamma_d$ limit tends to a saturation value,
$Q^{max}_{LT}=\frac{1}{2}T(1-T)V_b^2 2e^2/h$,
see the green curve in Fig.~\ref{fig:5}(d).
For example, $Q_{LT} \approx 0.1249 V_b^2 2e^2/h$ at $\Gamma_{d}=0.2$,
which is very close to $Q^{max}_{LT}\approx 0.125 V_b^2 2e^2/h$.
On the other hand, when the dissipation sources exist at both
the narrow constriction and the sample (in the red solid lines surrounding region in Fig.~\ref{fig:1}(a)),
the total local heat generation $Q_{LT}$ is much larger than that of
the dissipation sources in the green solid lines surrounding region,
but is less than $2Q^{max}_{LT}$.
In addition, our numerical results also indicate
that the total heat generation ($Q_T=Q_{LT}+\sum_{p=1}^6 Q_p$)
in the whole device is equal to the Joule heating $JV_{b}$,
since the electric current $J_{p}$ ($p\in\{{\bf i}, 2, 3, 5, 6\}$) is zero
and the energy of the system is conserved, demonstrating the validity of our numerical results.

\subsection{\label{B} Equivalent temperature}

\begin{figure}
	\includegraphics[scale=0.37]{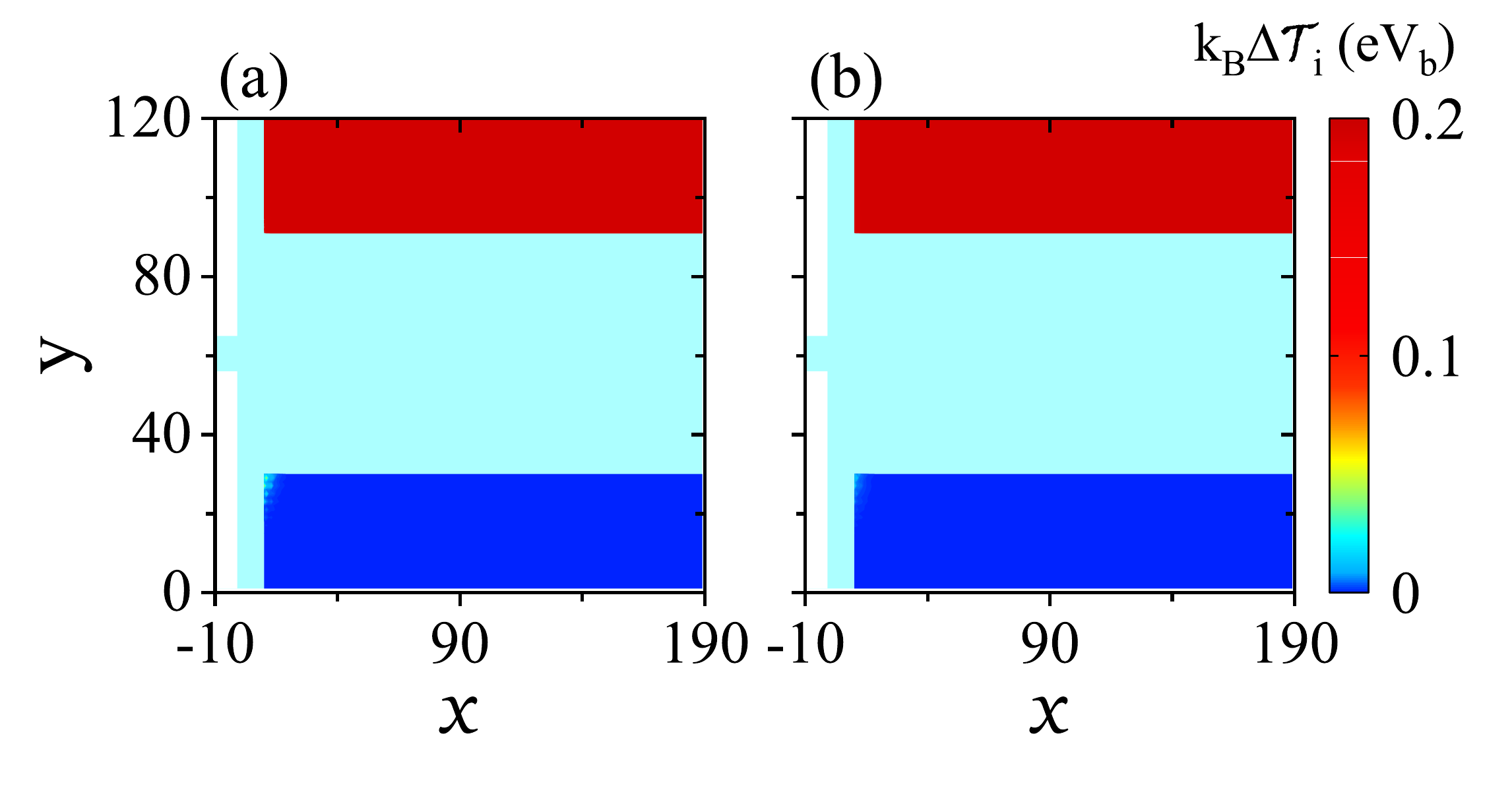}
	\centering
	\caption{Equivalent temperature $k_{B}\Delta\mathcal{T}_{\bf i}$
		($\Delta\mathcal{T}_{\bf i} \equiv \mathcal{T}_{\bf i}-\mathcal{T}$)
		vs lattice position ${\bf i}=(x,y)$ with $\Gamma_{d}=0.01$ (a) and $0.05$ (b).
		The background temperature $k_{B}\mathcal{T}/(eV_{b})=0.1$.
		The dissipation sources exist in the region surrounded
		by the green solid lines in Fig.~\ref{fig:1}(a).
		The light cyan area is added artificially to show the part without
		coupling of the virtual leads.}
	\label{fig:7}
\end{figure}

In the above, the temperature of the environment (the virtual
leads) is equal to the electronic system (the real leads).
While for poor thermal conductivity between the sample and the environment, the local heat generation $Q_{\bf i}$ disappears and the local electron temperature rises. Figures \ref{fig:6} and \ref{fig:7} show the equivalent temperature
$\Delta\mathcal{T}_{\bf i}$ ($\Delta\mathcal{T}_{\bf i} \equiv \mathcal{T}_{\bf i} -\mathcal{T}$)
versus the position ${\bf i}=(x,y)$ for the dissipation sources in the red
and green solid lines surrounding regions in Fig.~\ref{fig:1}(a), respectively.
The equivalent temperature along the downstream chiral flow direction (the upper edge of the system)
increases significantly because of the chiral heat transport in the QH plateau regime \cite{Nam,Granger}.
But at the upstream direction (the lower edge of the system),
the local temperature $\mathcal{T}_{\bf i}$
is almost equal to the background temperature $\mathcal{T}$.
What's more, the equivalent temperature does not reduce with the increase of the longitudinal location $x$
and is almost independent of the dissipation strength $\Gamma_{d}$.
For example, when the dissipation strength $\Gamma_{d}=0.005$,
$k_{B}\Delta\mathcal{T}_{\bf i}\approx0.1931eV_{b}$ at the upper edge,
which is almost the same as $k_{B}\Delta\mathcal{T}_{\bf i}\approx0.1926eV_b$ at $\Gamma_{d}=0.05$ [see Fig.~\ref{fig:6}]. Recent experimental work \cite{Halbertal1} has developed a temperature detection technique using an ultrasensitive scanning nano-thermometer
with a superconducting quantum interference device placed on a tip.
This technique allows one to obtain a spatial temperature variation
with submicrokelvin sensitivity.

\subsection{\label{C} Evolution of the energy distribution}

\begin{figure}
	\includegraphics[scale=0.31]{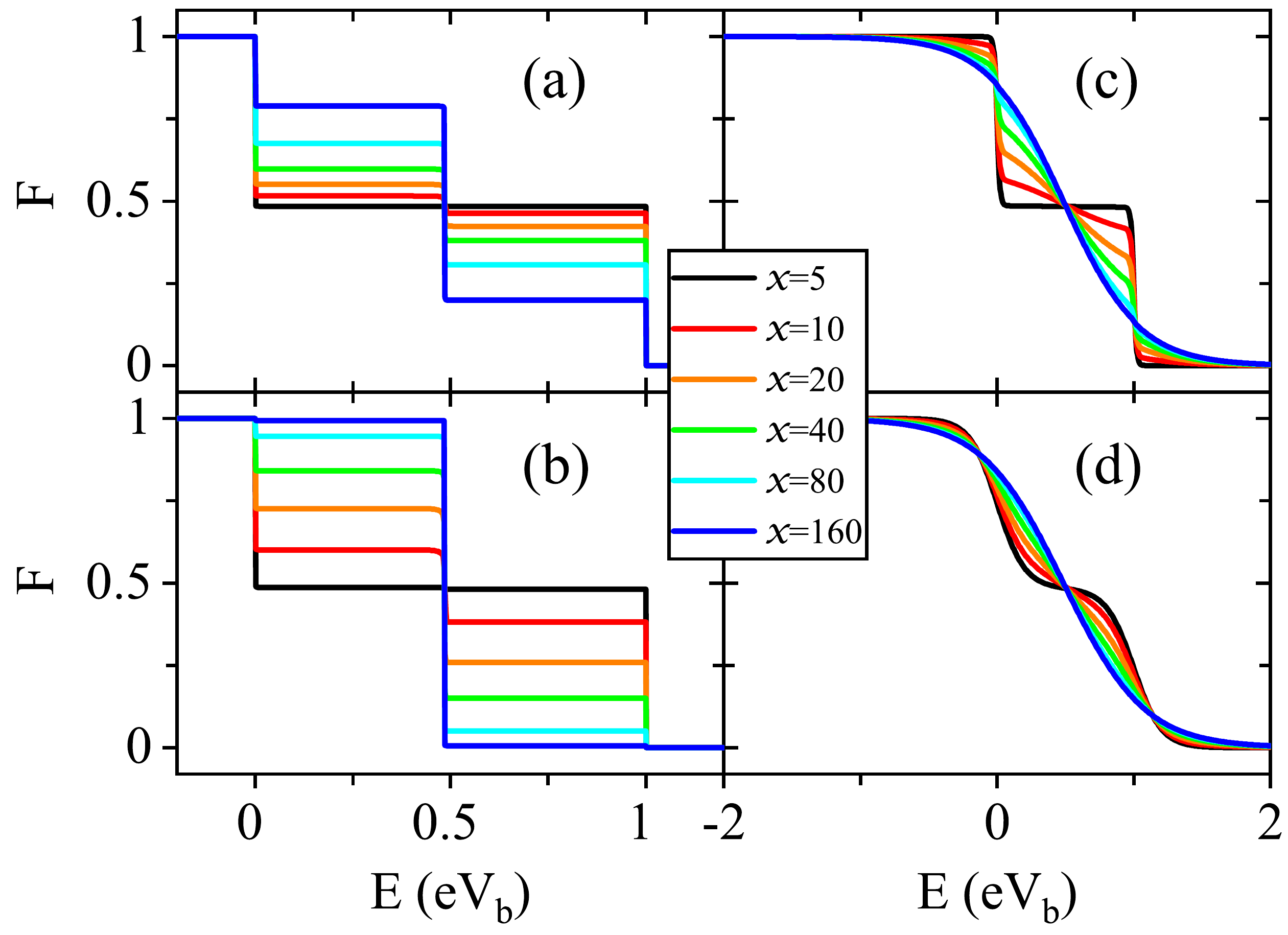}
	\centering
	\caption{
		(a-d) show distribution function $F$
		vs energy $E$ for different $x$ under fixed $y=115$.
		In (a) and (b), the temperature of the virtual leads are the same as the real leads,
		$\Gamma_{d}=0.01$ in (a) and 0.05 in (b).
		In (c) and (d), the virtual leads act as the temperature detecting terminals with $Q_{\bf i}=0$
		and $\Gamma_{d}=0.05$,
		$k_{B}\mathcal{T}/(eV_{b})$=0.01 in (c) and 0.1 in (d).
		The dissipation sources are in the green solid lines surrounding region.
	}
	\label {fig:8}
\end{figure}

In this subsection, let us discuss why thermal dissipation can occur
at the topologically protected chiral edge states in the QH plateau regime.
When the electrons injected from lead 1
arrive at the narrow constriction, some of them will be reflected back
and the others flow through the constriction,
leading to a non-integer transmission coefficient [see Fig.~\ref{fig:5}(c)].
Subsequently, the electron distribution at the downstream edge states
of the constriction is non-equilibrium.
For example, at zero temperature,
the distribution function of the downstream edge states satisfies $F(E)=1$ for
energy $E<0$, $F(E)=T(E)$ for $0<E<eV_b$, and $F(E)=0$ for $eV_b<E$
[see Figs.~\ref{fig:8}(a) and \ref{fig:8}(b)]. In other words,
the downstream edge states are completely occupied for $E<0$,
partially occupied for $0<E<eV_b$, and empty for $eV_b<E$.
In the presence of dissipation sources, these non-equilibrium electrons will
tend to be equilibrium and hence the thermal dissipation appears.
The maximum thermal dissipation is the energy difference between
the non-equilibrium states and the equilibrium ones, i.e.,
$Q^{max}_{LT}=\frac{1}{2}T(1-T)V_b^2 2e^2/h$.
At $B=0.02$, $T \approx 0.5$ and $Q^{max}_{LT} \approx 0.125 V_b^2 2e^2/h$.
Our numerical results show that $Q_{LT} \approx 0.1249 V_b^2 2e^2/h$ at $\Gamma_{d}=0.2$
[see Fig.~\ref{fig:5}(d)], which is very close to $Q^{max}_{LT}$.
Here are two points worth mentioning:
(i) Although the thermal dissipation occurs in the QH plateau regime,
backscattering cannot take place,
because during the dissipation process the electrons transfer from the high energy edge state
to the low one with unchanged propagating direction.
As a result, the QH plateaus and zero $R_{xx}$ can survive well [Figs.~\ref{fig:1}(b) and \ref{fig:1}(c)].
(ii) The thermal dissipation and entropy generation always occur
as long as the electron distribution is non-equilibrium,
no matter whether the system is topologically protected or not.
In a word, topology only protects no backscattering but cannot protect no dissipation.

Figures~\ref{fig:8}(a) and \ref{fig:8}(b) display the distribution $F_{\bf i}(E)$
at different locations $x$ along the downstream direction.
When the electrons pass through the constriction, they locate in
severe non-equilibrium states and the corresponding distribution
strongly deviates from the Fermi distribution function,
as can be seen from the curves of $x=5$ in Figs.~\ref{fig:8}(a) and \ref{fig:8}(b).
In the dissipation region, the local heat and entropy generate,
which is accompanied with the decrease of higher-energy electrons and the increase of lower-energy electrons,
and then the distribution function will evolve gradually from non-equilibrium to
equilibrium with increasing $x$.
After a long distance, the distribution will turn
back to the equilibrium Fermi distribution [Fig.~\ref{fig:8}(b)].
While for small $\Gamma_{d}$,
the distribution cannot return to the Fermi distribution even at $x=160$ [Fig.~\ref{fig:8}(a)].

Figures~\ref{fig:8}(c) and \ref{fig:8}(d) show the distribution $F_{\bf i}(E)$ versus energy $E$ for poor thermal conductivity between the sample and the environment. 
Similarly, the non-equilibrium distribution $F$ of the two-step shape
also evolves gradually into the equilibrium Fermi distribution with higher temperature, 
although the local heat generation is zero.
For small $x$ (near the narrow constriction),
the distribution of the electron is severely non-equilibrium with a two-step shape.
Along the +$x$ direction, the electron relaxation process occurs due to
the presence of the dissipation sources, and the non-equilibrium distribution
evolves gradually into the equilibrium Fermi distribution.
Without the energy loss, the temperature of the final equilibrium distribution rises. 
This indicates that dissipation and entropy can still increase.

\section{\label{secS4}CONCLUSIONS}
In summary, the thermal dissipation processes in the QH regime in graphene are studied. We find that the thermal dissipation can occur in the QH regime, with a relaxation length is affected by the dissipation strength. The thermal dissipation mainly appears along the downstream chiral flow direction of the constriction in the QH plateau regime, although the Hall plateaus and the zero longitudinal resistance remain well. While in the QH plateau transition regime, thermal dissipation mainly occurs in the bulk. Besides, for the poor thermal conductivity case, the local heat generation is zero but the local electron temperature rises, and it is not affected by the dissipation strength. Furthermore, accompanying with the thermal dissipation, the energy distribution of electrons evolves gradually
from non-equilibrium distribution to equilibrium Fermi distribution.
This work indicates that topology can only protect the propagating direction of carriers,
but cannot prohibit the emergence of dissipation and the increase of entropy.

\section*{ACKNOWLEDGMENTS}
This work was financially supported by National Key R and D Program of China (Grant No. 2017YFA0303301),
NSF-China (Grant No. 11921005 and No. 11874428), the Strategic Priority Research Program of Chinese
Academy of Sciences (Grant No. XDB28000000),
and Beijing Municipal Science \& Technology Commission (Grant No. Z191100007219013). We acknowledge the High-performance Computing Platform of Peking University for providing computational resources.

\section*{APPENDIX A: The phases $\phi_{\bf ij}$ in the longitudinal lead}

\begin{figure}
	\includegraphics[scale=0.2]{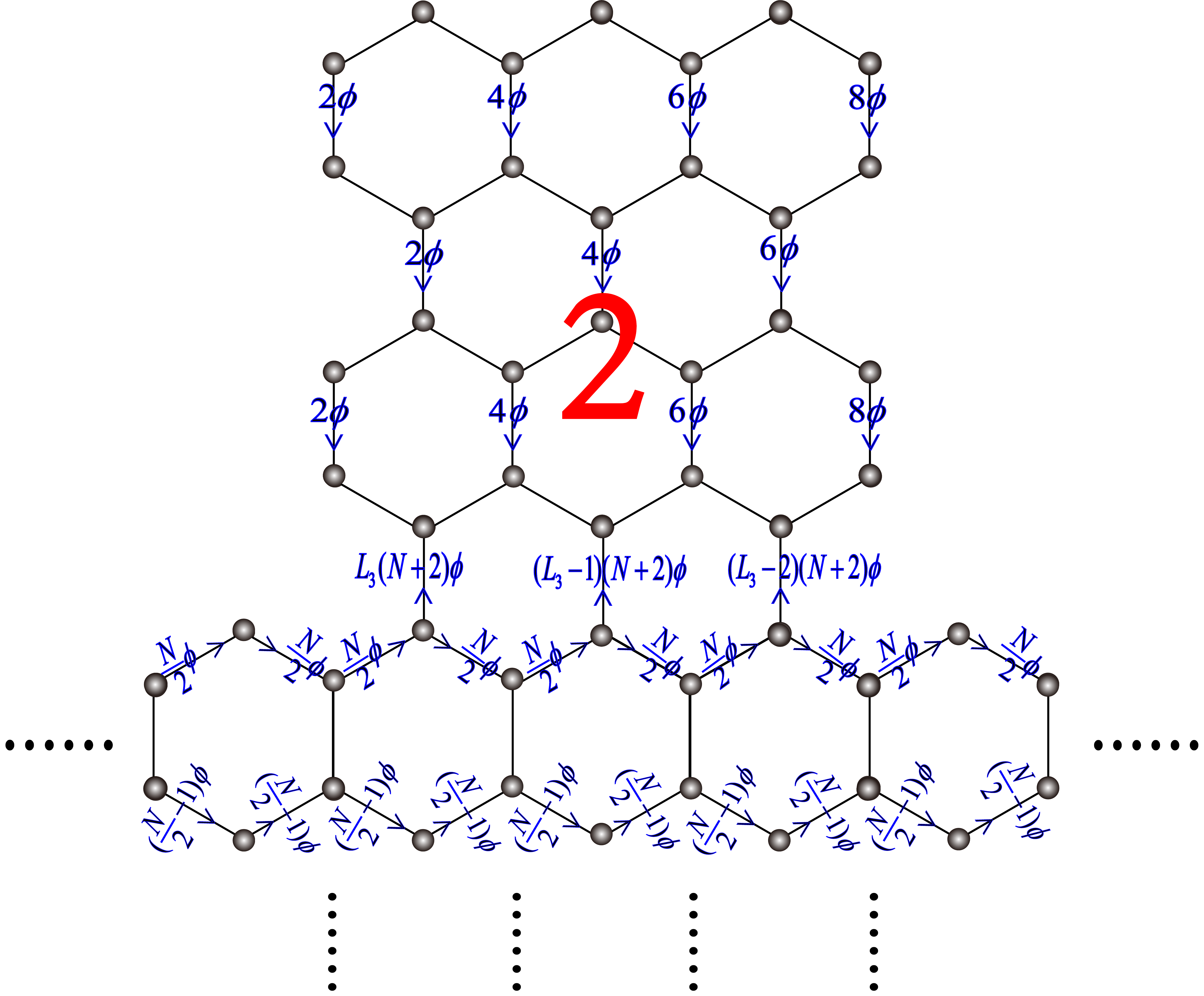}
	\centering
	\caption{
The zoomed-in figure of the lead 2 and the coupling between lead 2 and the central region in Fig.~\ref{fig:1}(a).
The arrows and values represent the phase $\phi_{\bf ij}$ from the site ${\bf i}$ to ${\bf j}$. The absence of arrow and value in some lines indicates that their phases are zero.
}
	\label {fig:9}
\end{figure}

In the numerical calculation, the Hamiltonian of the leads 2, 3, 5, 6 is required to obey
the translational invariance along $y$ direction, and
the Hamiltonian of the leads 1, 4 needs to obey the translational invariance along $x$ direction.
So we choose the vector potential $\vec{A}=(-By,0,0)$ in leads $1, 4$ and the central region,
$\vec{A}=(0,Bx,0)$ in the longitudinal leads 2, 3, 5, 6.
In Fig.~\ref{fig:9}, we show the phases $\phi_{\bf ij}$ in the lead 2 and the coupling between
lead 2 and the central region in detail, which are used in our calculation.
We can see that each hexagonal lattice has a phase of $2\phi$ along the clockwise direction,
which is equal to the number of the flux quantum $(3\sqrt{3}/2)a^2B/\phi_{0}$,
with $(3\sqrt{3}/2)a^2B$ the magnetic flux threading a single hexagon.
It is easy to prove that how to choose the phase $\phi_{\bf ij}$ has
no effect on the results, as long as the number of the flux quantum (the sum of the phase $\phi_{\bf ij}$ along the clockwise direction of the hexagonal lattice) in all hexagonal lattice is $2\phi$.
Choosing different phases $\phi_{\bf ij}$ is equivalent to
take the different gauge, so the results are the same.
In addition, for the leads 3, 5, 6 and their couplings to the central region,
we also guarantee that each hexagonal lattice has a phase of $2\phi$ along the clockwise direction
and the translational invariance along $y$ direction of leads 2, 3, 5, 6 is satisfied in our calculation.

\section*{APPENDIX B: The disappearance of the linear term of the voltage difference $\Delta V_{pq}$
in the expression of the heat generation}

\def\theequation{B\arabic{equation}}
\setcounter{equation}{0}

In Eq.~(\ref{eq:2}) in the main text,
$Q_{p}$ is electronic heat current from the central region to the lead $p$,
which includes the heat currents caused by the temperature difference,
by Peltier effect, and by the flow of electric current.
When a small bias is applied between leads 1 and 4,
the electric current flows through the device,
then the thermal dissipation occurs and the Joule heat generates
while in the presence of the dissipation sources~\cite{aref11,aref12}.
Notice that the current-induced local heat generation is proportional to $(\Delta V_{pq})^2$
($\Delta V_{pq} \equiv V_p-V_q$).
This is essentially different from the heat current caused
by Peltier effect, which is proportional to $\Delta V_{pq}$.

Now let us discuss under what conditions the current-induced local heat generation
(the quadratic term $(\Delta V_{pq})^2$) dominates the heat current $Q_p$.
In order to clearly show this issue,
we set that the temperatures of all leads are equal ($\mathcal{T}_p =\mathcal{T}$).
Expanding the Fermi function up to the second order term,
we have:
\begin{eqnarray} \label{eq:B1}
    f_p(E) \approx f(E)-eV_p f'(E) +\frac{e^2V_p^2}{2} f''(E)
\end{eqnarray}
where $f(E)=\frac{1}{e^{\epsilon}+1}$,
$ f'(E)=- \frac{1}{k_B\mathcal{T}}
\frac{e^{\epsilon}}{(e^{\epsilon}+1)^2}$,
and $f''(E)=\frac{1}{k_B^2 \mathcal{T}^2}
\frac{e^{\epsilon} - e^{-\epsilon}}
{(e^{\epsilon}+1)^2 (e^{-\epsilon}+1)^2}$ with $\epsilon =\frac{E-E_F}{k_B\mathcal{T}}$.
Considering the Sommerfeld expansion for the transmission coefficient:
\begin{eqnarray} \label{eq:B2}
	T_{pq}(E)=T_{pq}(E_F)+(E-E_F)T'_{pq},
\end{eqnarray}
with $T'_{pq} =\left.\frac{dT_{pq}(E)}{dE}\right|_{E=E_F}$.
Substituting Eqs.~(\ref{eq:B1}) and (\ref{eq:B2}) into Eq.~(\ref{eq:2}),
$Q_p$ changes into:
\begin{eqnarray} \label{eq:B3}
    Q_{p}&=&\frac{2}{h}\sum_{q}\int dE (E-\mu_p)
    \left[T_{pq}(E_F)+(E-E_F)T'_{pq}\right]\times \nonumber\\
       & &  \left[e(V_p-V_q) f'(E) -(e^2/2)(V_p^2-V_q^2) f''(E)\right]  \nonumber\\
   &=& \frac{2}{h} \sum_{q}
   \left\{ -\frac{\pi^2}{3} k_B^2\mathcal{T}^2 e\Delta V_{pq}
 T'_{pq} + \frac{1}{2} e^2(\Delta V_{pq})^2 T_{pq}(E_F) \right.\nonumber\\
    & &  + \left.\frac{1}{2}e^3 V_p(V^2_p-V^2_q) T'_{pq}
    \right\}
\end{eqnarray}
At the small voltage difference $\Delta V_{pq}$ limit and the finite temperature $\mathcal{T}$,
the leading term of $Q_p$ is $-\frac{\pi^2}{3} k_B^2\mathcal{T}^2 e\Delta V_{pq} T'_{pq}$,
which is the linear term $\Delta V_{pq}$, depends on $T'_{pq}(E_F)$ (not $T_{pq}(E_F)$),
and describes Peltier effect~\cite{addTE1,addTE2}.
However, when both the voltage difference $\Delta V_{pq}$
and the temperature $\mathcal{T}$ are small,
the leading term of $Q_p$ in Eq.~(\ref{eq:B3}) is the second term, $\frac{1}{2} (e\Delta V_{pq})^2 T_{pq}(E_F)$.
This term is proportional to $(\Delta V_{pq})^2$,
depends on $T_{pq}(E_F)$, and describes the Joule heat generation by the electric current.
Because the present work studies the heat generation and assumes the small temperature
and small voltage, the term $\frac{1}{2} (e\Delta V_{pq})^2 T_{pq}(E_F)$ emerges
in Eqs.~(\ref{eq:4}) and (\ref{eq:5}) in the main text.

\end{document}